\documentclass[lettersize,journal]{IEEEtran}
\usepackage{amsmath,amsfonts}
\usepackage{algorithmic}
\usepackage{algorithm}
\usepackage{array}
\usepackage[caption=false,font=footnotesize,labelfont=sf,textfont=sf]{subfig}
\usepackage{textcomp}
\usepackage{stfloats}
\usepackage{url}
\usepackage{verbatim}
\usepackage{graphicx}
\usepackage{cite}
\usepackage{booktabs}
\usepackage{multicol}
\usepackage{graphicx}
\usepackage[export]{adjustbox}
\usepackage{array}
\usepackage{xcolor}
\newcolumntype{P}[1]{>{\centering\arraybackslash}p{#1}}
\usepackage{amssymb}

\usepackage{helvet} 

\usepackage{setspace} 

\usepackage{tikz}
\usetikzlibrary{shapes,arrows,positioning,calc}

\usepackage{afterpage}

\tikzset{
  block/.style = {rectangle, draw, fill=white, text centered, rounded corners, minimum height=2em, minimum width=1em}, 
  block2/.style = {rectangle, draw, fill=white, text centered, rounded corners, minimum height=2em, minimum width=1em},
  input/.style = {coordinate}, 
  output/.style = {coordinate},
  sum/.style = {draw, circle, inner sep=0pt, minimum size=0.5cm}, 
  sum2/.style = {draw, circle, inner sep=0pt, minimum size=0.5cm},
  arrow/.style = {->, thick, >=latex} 
}

\usepackage{hyperref}

\usepackage{bm}

\begin{document}

\bstctlcite{IEEEexample:BSTcontrol} 

\title{Digital Particle Communication under Flow: Frequency Analysis and Pulse Design}

\author{Fatih~Merdan,
       Ozgur B.~Akan,~\IEEEmembership{Fellow,~IEEE}

\thanks{Fatih Merdan and O.B. Akan are with the Centre for neXt
Communications (CXC), Department of Electrical and Electronics Engineering, Koc¸ University, Istanbul 34450, Türkiye (e-mail:  fmerdan25@ku.edu.tr, akan@ku.edu.tr).}
\thanks{O. B. Akan is also with the Centre for neXt Communications (CXC), Department of Engineering, University of Cambridge, Cambridge CB3 0FA, United Kingdom (e-mail: oba21@cam.ac.uk).}}

\markboth{}%
{}

\maketitle

\begin{abstract}
In particle communication, particles released for different symbols can reach the receiver at overlapping times, making reliable detection difficult. This paper develops a digital communication framework that connects particle transport, pulse shaping, and receiver processing for three-dimensional free-space diffusion-advection channels. We derive the continuous-time and discrete-time frequency responses and explain how flow speed affects channel delay and bandwidth. Under advection-dominated conditions, the group delay approaches $R/V$, and the one-sided $3$-dB bandwidth scales as $V^{3/2}$. We design nonnegative, unit-energy transmit pulses using a packet-level SINR criterion that includes interference from both preceding and subsequent symbols under additive white Gaussian receiver noise. Simulations validate the analytical approximations and show that optimized pulses spread across more of the symbol interval when noise is substantial and become narrower as noise decreases. BER evaluations across SNR and symbol duration compare optimized, rectangular, and single-impulse pulses with several detection methods, identifying the operating conditions in which transmit-pulse design improves reliability.
\end{abstract}

\begin{IEEEkeywords}
Channel models, digital communication, frequency response, intersymbol interference, molecular communication, pulse shaping methods.
\end{IEEEkeywords}

\section{Introduction}
\IEEEPARstart{M}{olecular} communication (MC) conveys information through the release and transport of molecules or particles. Its applications span microscopic and macroscopic environments, with propagation governed by diffusion, advection, or their combination \cite{Channel_Modeling_for_Diffusive_Molecular_CommunicationA_Tutorial_Review}. Although transmitter and receiver implementations differ across scales, temporal dispersion is a common challenge: particles released during one symbol interval can affect observations associated with other symbols, producing intersymbol interference (ISI). Understanding how transport parameters determine channel memory, and how pulse shaping and detection exploit this structure, is therefore essential for reliable communication. Receiver design has received considerable attention in MC. Likelihood-based detection, sequence estimation, and equalization have been developed for diffusion channels \cite{Receiver_Design_for_Molecular_Communication,On_Receiver_Design_for_DiffusionBased_Molecular_Communication}, while ligand--receptor observations have motivated receiver-specific ML methods \cite{Maximum_Likelihood_Detection_With_Ligand_Receptors_for_DiffusionBased_Molecular_Communications_in_Internet_of_BioNano_Things}. Transport with drift or a moving propagation medium has also been investigated \cite{Molecular_Communication_Using_Brownian_Motion_With_Drift,Optimal_Receiver_Design_for_Diffusive_Molecular_Communication_With_Flow_and_Additive_Noise,NanoMachine_Molecular_Communication_over_a_Moving_Propagation_Medium}. These studies establish important channel and receiver models. Building on this foundation, the present work develops a unified discrete-time representation of the complete particle communication link, connecting physical transport, transmit-pulse shaping, receiver filtering, and symbol detection. This formulation enables the system to be analyzed and designed from a communication-theoretic perspective using established digital communication tools.

Frequency-domain approaches have been used to characterize molecular transport in microfluidic flow channels \cite{Molecular_Transport_in_Microfluidic_Channels_for_FlowInduced_Molecular_Communication}, model propagation and receiver dynamics in a microfluidic system with a graphene bioFET receiver \cite{FrequencyDomain_Model_of_Microfluidic_Molecular_Communication_Channels_With_Graphene_BioFETBased_Receivers}, and develop equalization methods for diffusive MC \cite{Frequency_Domain_Analysis_and_Equalization_for_Molecular_Communication}. To connect such approaches to particle communication, it is useful to distinguish the temporal variation of the information-bearing signal from the oscillation of a carrier. In conventional digital communication, symbols form a time-varying baseband waveform that can subsequently modulate an oscillatory carrier. Its baseband frequency content is determined by the symbol rate and pulse shape. In a particle communication system, the release waveform itself carries the information, with no mechanism for translating it onto a separate oscillatory carrier. Nevertheless, the temporal signaling structure is preserved: symbols are transmitted at regularly spaced intervals, and the corresponding release waveform produces a time-varying concentration at the receiver. Thus, frequency characterizes how rapidly the release and concentration signals vary, while the channel frequency response describes how transport attenuates and delays these variations. Building on the existing frequency-domain studies, this work establishes explicit connections between physical transport parameters and digital communication quantities for a three-dimensional free-space diffusion-advection channel. We derive the magnitude, phase, and group-delay responses and relate the effective bandwidth to flow speed. In the advection-dominated regime, the resulting approximations identify a nearly constant delay $R/V$ and a one-sided $3$-dB bandwidth proportional to $V^{3/2}$. These relations provide a physical basis for interpreting channel memory and selecting symbol durations for pulse design and detection.

Pulse shaping provides a complementary approach to controlling dispersion. Existing work includes SINR-based receiver-filter design for molecule-counting observations \cite{On_the_Design_of_Matched_Filters_for_Molecule_Counting_Receivers}, energy-constrained transmit signal shaping \cite{SchedulingBased_Transmit_Signal_Shaping_in_EnergyConstrained_Molecular_Communications}, and symbol-dependent transmission profiles that improve response separability while matching interference tails \cite{Optimal_Transmission_Sequence_Design_with_ISI_Matching_in_Molecular_Communication}. Most of the receiver-design and ISI-mitigation studies describe ISI through residual contributions from previous releases, without explicitly identifying precursor interference relative to a delayed symbol-decision reference \cite{Receiver_Design_for_Molecular_Communication,On_Receiver_Design_for_DiffusionBased_Molecular_Communication,Optimal_Receiver_Design_for_Diffusive_Molecular_Communication_With_Flow_and_Additive_Noise,ISI_Mitigation_Techniques_in_Molecular_Communication}. However, particle transport can produce both precursor and postcursor contributions at such a reference. Particles released together do not all reach the receiver at the same time: some arrive earlier, while others arrive later. When the receiver observes the concentration to detect a particular symbol, particles released for previous symbols may still contribute, producing postcursor ISI. Because detection occurs after a propagation delay, subsequent symbols may also have been transmitted. Some particles from these subsequent releases can reach the receiver by the same observation time, producing precursor ISI. Both effects are represented in the mean concentration response and exist independently of receiver filtering. The sampling reference and receiver filter determine their corresponding coefficients in the symbol-rate model. Our formulation designs a common nonnegative, unit-energy transmit pulse using a packet-level SINR criterion that accounts for desired-signal gain, receiver noise, and both precursor and postcursor ISI. The resulting pulse shapes illustrate how the balance between noise and ISI influences the preferred temporal concentration of the release waveform. We compare these optimized pulses with rectangular and single-impulse pulses using established detection and equalization methods. BER results across SNR and symbol duration demonstrate the interaction between pulse shaping, channel memory, and receiver processing.

The remainder of this paper is organized as follows. Section~II presents the time-varying diffusion-advection channel model and its discrete-time representation. Section~III develops the frequency-domain analysis for constant flow. Section~IV presents the digital communication system, pulse-design formulation, and detection methods. Section~V reports the simulation results, and Section~VI concludes the paper.

\section{Channel Discretization}

The mass-based communication system considered in this paper consists of a transmitter that releases particles, a diffusion-advection channel, and a receiver. Such systems can operate at microscopic or macroscopic scales. The received signal is commonly represented by particle counts in microscopic systems and by concentration in macroscopic systems. However, concentration-based descriptions are also used in microscopic systems when the number of particles is sufficiently large \cite{A_Comprehensive_Survey_of_Recent_Advancements_in_Molecular_Communication}. Practical transmitter and receiver implementations differ between these scales, and microscopic devices may be constrained in terms of sampling frequency and computational resources. In this work, we abstract these implementation constraints and assume devices with sufficient sampling and computational capabilities to develop a theoretical framework that can subsequently be specialized to either scale. Specifically, the transmitter digitally generates pulses and controls the particle release rate accordingly, while the receiver samples the concentration signal and processes the samples digitally. We further assume that receiver electronic noise is the dominant noise contribution and represent it by additive white Gaussian noise (AWGN). This approximation is motivated by thermal noise in the receiver circuitry and provides a tractable basis for the communication-theoretic analysis.

Particle transport under diffusion and advection is described by the diffusion-advection equation. For $\nabla\cdot\mathbf{v}(\mathbf{x},t)=0$, this equation takes the form
\begin{equation}
\label{eq:pde_c}
\begin{split}
\frac{\partial c}{\partial t}(\mathbf{x},t)= D\nabla^{2}c(\mathbf{x},t) - \mathbf{v}(\mathbf{x},t)\cdot\nabla c(\mathbf{x},t)
+ S(\mathbf{x},t),
\end{split}
\end{equation}
where $c(\mathbf{x},t)$ denotes the particle concentration at position $\mathbf{x}$ and time $t$, $D$ is the diffusion coefficient, $\mathbf{v}(\mathbf{x},t)$ is the flow velocity vector, and $S(\mathbf{x},t)$ is the source term. The operators $\nabla$ and $\nabla^{2}$ denote the spatial gradient and Laplacian, respectively. The propagation medium may be a liquid or a gas. The diffusion coefficient depends on the particle and medium properties, while the flow field depends on the geometry and operating conditions. Flow characteristics also differ across applications: microscopic systems often involve laminar flow, whereas macroscopic systems may exhibit turbulent flow \cite{A_Comprehensive_Survey_of_Recent_Advancements_in_Molecular_Communication}. Within the adopted transport model, these differences are represented through the diffusion coefficient and the velocity field. This work therefore examines the channel from a communication-theoretic perspective, using a common mathematical framework rather than specializing the analysis to a particular scale or application.

We will follow the channel impulse response derivation steps given in \cite{Airborne_Particle_Communication_Through_Timevarying_DiffusionAdvection_Channels}, but with three flow components in free space. Here, it is assumed that the flow is the same in all space but varying with time. Assuming a point source at position \(\mathbf{z}_{0}\) and zero initial concentration for the particle, the channel impulse response (CIR) is written as
\begingroup
\small
\begin{align}
h(\tau,t)
=
\frac{u(\tau)}{(4\pi D\tau)^{3/2}}
\exp\!\left[
\frac{-1}{4D\tau}
\sum_{i=1}^{3}
\left(
x_i-z_{0i}
-
\int_{t-\tau}^{t}v_i(\xi)\,d\xi
\right)^{2}
\right],
\label{eq:time_varying_impulse_response}
\end{align}
\endgroup
\noindent where \(u(\tau)\) is the unit step function. Let $f_s$ denote the digital sampling rate and $T_s \triangleq 1/f_s$ the sampling period. Discrete-time signals are indexed by $n\in\mathbb{Z}_{\ge 0}$ with $t=nT_s$. Let $T_{\mathrm{sym}}$ denote the symbol duration and $N_p \triangleq \frac{T_{\mathrm{sym}}}{T_s}$ the number of digital samples per symbol interval. We consider linearly modulated signals with pulses $p_\ell[n]$ having disjoint supports within a symbol duration, which ensures orthogonality: $\sum_{n=0}^{N_p-1} p_k[n]\,p_\ell[n]=0,\ k\neq \ell$. For each symbol index $m$, define the discrete-time symbol waveform $x_m[n] \triangleq \sum_{\ell=1}^{N} s_{m,\ell}\,p_{\ell}[n], 0\le n \le N_p-1,$ where $N$ is the modulation dimension, $s_{m,\ell}$ is the weight of the $\ell$th pulse in the $m$th symbol. The overall transmit sequence is obtained as $q[n]=\sum_{m} x_m[n-mN_p] =\sum_{m}\sum_{\ell=1}^{N} s_{m,\ell}\,p_{\ell}[n-mN_p]$. We use the one-sided definition of the rectangular function where
\begin{align}
\mathrm{rect}(x)\triangleq
\begin{cases}
1, & 0\le x<1,\\
0, & \text{otherwise}.
\end{cases}
\label{eq:rect_defn}
\end{align}
Under zero-order hold (ZOH) reconstruction, the continuous-time transmit waveform $q(t)$ is generated from the discrete-time sequence $q[n]$ as $q(t)=\sum_{k\in\mathbb{Z}} q[k]\;\mathrm{rect}\!\left(\frac{t-kT_s}{T_s}\right)$. The received concentration is obtained after the linear time-varying (LTV) channel as $c(t)=\int_{0}^{t} h(\tau,t)\,q(t-\tau)\,d\tau$. The resulting sampled channel output is $c[n]\triangleq c(nT_s)=\int_{0}^{nT_s} h(\tau,nT_s)\,q(nT_s-\tau)\,d\tau$. Substituting the ZOH expansion $q(nT_s-\tau)=\sum_{k\in\mathbb{Z}} q[k]\;\mathrm{rect}\!\left(\frac{(nT_s-\tau)-kT_s}{T_s}\right)$ yields $c[n]=\sum_{k\in\mathbb{Z}} q[k]\int_{0}^{nT_s} h(\tau,nT_s)\, \mathrm{rect}\!\left(\frac{nT_s-\tau-kT_s}{T_s}\right)\,d\tau$. Hence $c[n]
=\sum_{k\in\mathbb{Z}} q[k] \int_{\max\{0,(n-k-1)T_s\}}^{\min\{nT_s,(n-k)T_s\}} h(\tau,nT_s)\,d\tau$ is obtained. Only indices $k\in\{0,1,\dots,n-1\}$ yield non-empty integration intervals, and therefore $c[n]=\sum_{k=0}^{n-1} q[k]\int_{(n-k-1)T_s}^{(n-k)T_s} h(\tau,nT_s)\,d\tau$ is written. We define the exact discrete-time channel coefficients as
\begin{align}
f[n,k]\triangleq \int_{(n-k-1)T_s}^{(n-k)T_s} h(\tau,nT_s)\,d\tau,
\label{eq:DT_channel_coeffs}
\end{align}
which gives the discrete-time input-output relation $c[n]=\sum_{k=0}^{n-1} f[n,k]\;q[k]$. We next analyze the channel frequency response and design transmit pulses using digital communication techniques. For this purpose, we consider constant flow, under which the channel is linear time-invariant (LTI). For constant flow, the flow velocity is written as $\mathbf{v} = (v_1,v_2,v_3)$. Therefore, $\int_{t-\tau}^{t} v_i(\xi)\,d\xi = v_i \tau$ is obtained.

\section{Frequency Domain Analysis of the Constant Flow LTI Channel}

Defining $r_i \triangleq x_i-z_{0i}$, the transmitter-receiver distance is written as $R\triangleq\sqrt{r_1^2+r_2^2+r_3^2}$. Moreover, the magnitude of the constant flow is defined as $V\triangleq\sqrt{v_1^2+v_2^2+v_3^2}$. Using these, the CIR is written as 
\begingroup
\small
\begin{equation}
\label{eq:factored_constant_flow_cir}
\begin{split}
h(\tau)=
\frac{u(\tau)\tau^{-3/2}}{(4\pi D)^{3/2}}
\exp\!\left(\frac{\sum_{i=1}^{3}r_i v_i}{2D}\right)
\exp\!\left(-\frac{R^2}{4D\tau}-\frac{V^2\tau}{4D}\right).
\end{split}
\end{equation}
\endgroup
The continuous-time frequency response is $H(f)\triangleq\int_{-\infty}^{\infty}h(\tau)e^{-j2\pi f\tau}\,d\tau$. Therefore, the frequency response is found as
\begin{equation}
\label{eq:frequency_response_with_integral}
\begin{split}
H(f)
&=
\frac{1}{(4\pi D)^{3/2}}
\exp\left(
\frac{r_1v_1+r_2v_2+r_3v_3}{2D}
\right)
I(R,f),
\end{split}
\end{equation}
\noindent where $I(R,f)\triangleq\int_{0}^{\infty}\tau^{-3/2}\exp\left[-\frac{R^2}{4D\tau}-\left(\frac{V^2}{4D}+j2\pi f\right)\tau\right]d\tau$. Using Eq. 3.471.9 and Eq. 8.469.3 of \cite{Table_of_Integrals_Series_and_Products}, $I=\frac{2\sqrt{\pi D}}{R}\exp\left[-\frac{R}{2D}\sqrt{V^2+j8\pi Df}\right]$ is obtained. Therefore, the frequency response of the constant-flow channel is
\begingroup
\small
\begin{equation}
\label{eq:free_space_frequency_response}
\begin{split}
H(f)
&=
\frac{1}{4\pi DR}
\exp\left[
\frac{r_1v_1+r_2v_2+r_3v_3}{2D}
-\frac{R}{2D}
\sqrt{V^2+j8\pi Df}
\right].
\end{split}
\end{equation}
\endgroup
Next, we calculate the magnitude and phase responses of the LTI channel. To do this we define $a\triangleq\frac{V^2}{4D^2}$ and $b\triangleq\frac{2\pi f}{D}$. Then $\sqrt{a+jb}\triangleq\rho+j\eta$ is written where the principal square-root branch is used. The real and imaginary parts are $\rho=\sqrt{\frac{\sqrt{a^2+b^2}+a}{2}}$ and $\eta(f)=\operatorname{sgn}(f)\sqrt{\frac{\sqrt{a^2+b^2}-a}{2}}$. Defining $A\triangleq\frac{1}{4\pi D}\exp\left(\frac{r_1v_1+r_2v_2+r_3v_3}{2D}\right)$ and $G(f)\triangleq\frac{\exp(-R\rho)}{R}$, the frequency response can be written as $H(f)=A\,G(f)\exp(-jR\eta) = A\,G(f)\left[\cos(R\eta)-j\sin(R\eta)\right]$. The real and imaginary parts of the frequency response are
\begin{equation}
\label{eq:frequency_response_real_imaginary}
\begin{split}
\operatorname{Re}\{H(f)\}
&=
A\,G(f)\cos(R\eta), \\
\operatorname{Im}\{H(f)\}
&=
-A\,G(f)\sin(R\eta).
\end{split}
\end{equation}
The magnitude response is $|H(f)|=A\,G(f)$ which can be written as
\begingroup
\small
\begin{equation}
\label{eq:magnitude_response_final}
\begin{split}
|H(f)|
&=
\frac{1}{4\pi DR}
\exp\left(
\frac{\sum_{i=1}^{3}r_i v_i}{2D}
\right)
\exp\left[
-R\sqrt{\frac{\sqrt{a^2+b^2}+a}{2}}
\right].
\end{split}
\end{equation}
\endgroup
The unwrapped phase response, chosen to be zero at $f=0$, is $\angle H(f)=-R\eta(f)$. Substituting the expression for \(\eta\) yields
\begin{equation}
\label{eq:phase_response_final}
\begin{split}
\angle H(f)
&=
-R\operatorname{sgn}(f)
\sqrt{
\frac{\sqrt{a^2+b^2}-a}{2}
}.
\end{split}
\end{equation}
The group delay is defined as $\tau_g(f)\triangleq-\frac{1}{2\pi}\frac{d\gamma(f)}{df}$ where \(\gamma(f)\) denotes the unwrapped phase response. Calculating this expression yields
\begin{equation}
\label{eq:exact_group_delay}
\begin{split}
\tau_g(f)
&=
\frac{R}{2D}
\frac{
\sqrt{
\dfrac{\sqrt{a^2+b^2}+a}{2}
}
}{
\sqrt{a^2+b^2}
}.
\end{split}
\end{equation}
The advection-dominated approximation is based on $\left|\frac{b}{a}\right|\ll1$ which corresponds to $\frac{8\pi D|f|}{V^2}\ll1$. Under this condition, $\sqrt{1+\frac{b^2}{a^2}}\approx(1+\frac{b^2}{2a^2})$ can be used and the resulting group delay is found as
\begin{equation}
\label{eq:constant_group_delay_approximation}
\begin{split}
\tau_g(f)
\approx
\frac{R}{V}.
\end{split}
\end{equation}
Thus, over the frequency range satisfying this condition, the phase response is approximately linear and the group delay is approximately constant, equal to the advective transit time $R/V$. We also find an approximate relation between effective bandwidth and flow speed. For this, we normalize the magnitude response with $|H(0)|=\frac{1}{4\pi DR}\exp(\frac{r_1v_1+r_2v_2+r_3v_3}{2D})\exp(-\frac{RV}{2D})$ to obtain
\begin{equation}
\label{eq:normalized_magnitude_response}
\begin{split}
\frac{|H(f)|}{|H(0)|}
&=
\exp\left[
-R\left(
\rho(f)-\frac{V}{2D}
\right)
\right].
\end{split}
\end{equation}
Under the same advection-dominated condition \(\left|b/a\right|\ll1\), we obtain
\begin{equation}
\label{eq:normalized_magnitude_gaussian}
\begin{split}
\frac{|H(f)|}{|H(0)|}
\approx
\exp\left(
-\frac{4\pi^2RD}{V^3}f^2
\right).
\end{split}
\end{equation}
We define the one-sided effective bandwidth \(f_B(\alpha)\) through the magnitude threshold $\frac{|H(f_B)|}{|H(0)|}=\alpha,$ where $0<\alpha<1$. This results in $f_B(\alpha)\approx\sqrt{-\frac{V^3\ln\alpha}{4\pi^2RD}}$. Setting $\alpha=1/\sqrt{2}$ for the conventional $-3$ dB threshold gives
\begin{equation}
\label{eq:three_db_bandwidth}
\begin{split}
f_{\mathrm{3dB}}
\approx
\sqrt{
\frac{V^3\ln2}{8\pi^2RD}
}.
\end{split}
\end{equation}
Hence, for fixed $R$ and $D$, the effective bandwidth scales approximately as $f_{\mathrm{3dB}}\propto V^{3/2}$, indicating a superlinear increase with flow speed under the advection-dominated condition.

Under constant flow, the discrete channel coefficients given in \eqref{eq:DT_channel_coeffs} reduces to $h[k]\triangleq \int_{(k-1)T_s}^{(k)T_s} h(\tau)\,d\tau$ for $k>0$ and $h[k]=0$ for $k\leq0$. The sampling operation is expressed as $h[k]=\bar h(kT_s)$, where the continuous-time signal is $\bar h(t)=h(t)*\operatorname{rect}(\frac{t}{T_s})=\int_{t-T_s}^{t}h(\tau)\,d\tau$. Using the normalized $\operatorname{sinc}$ function $\operatorname{sinc}(x)=\frac{\sin(\pi x)}{\pi x}$, the continuous-time Fourier transform of the rectangular function is written as $\mathcal{F}\{\operatorname{rect}(\frac{t}{T_s})\}=T_se^{-j\pi fT_s}\operatorname{sinc}(fT_s)$. Consequently, we obtain
\begin{equation}
\bar H(f)
=
H(f)
T_s
e^{-j\pi fT_s}
\operatorname{sinc}(fT_s).
\label{eq_H_bar}
\end{equation}
The discrete-time Fourier transform of \(h[k]\) is defined as $H_d(e^{j\omega})=\sum_{k=-\infty}^{\infty}h[k]e^{-j\omega k}$. Since \(h[k]\) is obtained by uniformly sampling \(\bar h(t)\), the exact continuous-to-discrete frequency-domain relation is $H_d(e^{j\omega})=\frac{1}{T_s}\sum_{r=-\infty}^{\infty}\bar H(\frac{\omega-2\pi r}{2\pi T_s})$, which yields
\begingroup
\small
\begin{equation}
\begin{aligned}
H_d\left(e^{j\omega}\right)
={}&
\sum_{r=-\infty}^{\infty}
H
\left(
\frac{\omega-2\pi r}{2\pi T_s}
\right)
e^{-j(\omega-2\pi r)/2}
\operatorname{sinc}
\left(
\frac{\omega-2\pi r}{2\pi}
\right).
\end{aligned}
\label{eq_exact_discrete_frequency_response}
\end{equation}
\endgroup
If the shifted copies of \(\bar H(f)\) do not overlap significantly, the aliasing terms with \(r\neq0\) can be neglected over the principal frequency interval \(-\pi\leq\omega\leq\pi\). In this case,
\begin{equation}
H_d\left(e^{j\omega}\right)
\approx
H
\left(
\frac{\omega}{2\pi T_s}
\right)
e^{-j\omega/2}
\operatorname{sinc}
\left(
\frac{\omega}{2\pi}
\right),
-\pi\leq\omega\leq\pi.
\label{eq_no_alias_discrete_frequency_response}
\end{equation}
The complete approximate DTFT is obtained by periodically extending \eqref{eq_no_alias_discrete_frequency_response} with period \(2\pi\). Since $\operatorname{sinc}(\frac{\omega}{2\pi})>0, -\pi\leq\omega\leq\pi$, the approximate magnitude response is
\begin{equation}
\left|
H_d\left(e^{j\omega}\right)
\right|
\approx
\left|
H
\left(
\frac{\omega}{2\pi T_s}
\right)
\right|
\operatorname{sinc}
\left(
\frac{\omega}{2\pi}
\right).
\label{eq_discrete_magnitude_response}
\end{equation}
Similarly, the approximate phase response is
\begin{equation}
\angle
H_d\left(e^{j\omega}\right)
\approx
\angle
H
\left(
\frac{\omega}{2\pi T_s}
\right)
-
\frac{\omega}{2}.
\label{eq_discrete_phase_response}
\end{equation}

\begin{figure}[!h]
  \centering
  \includegraphics[width=1\linewidth]{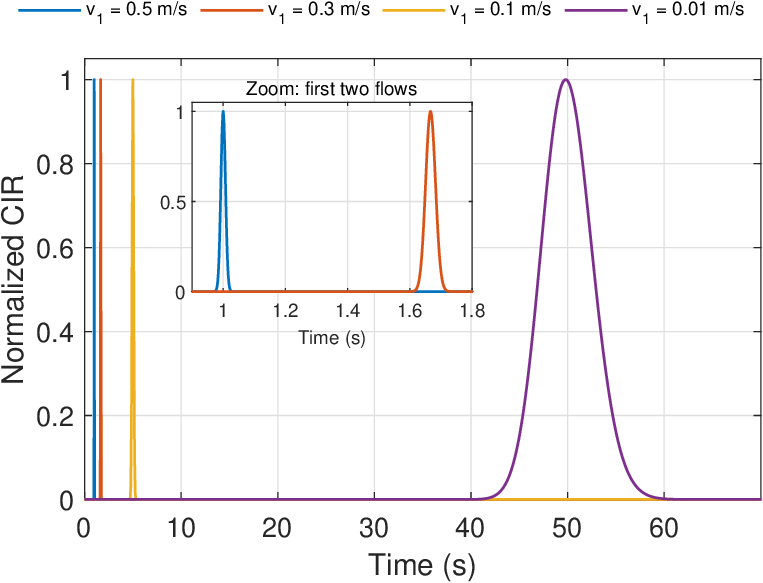}
  \caption{Normalized CIRs for different constant flow speeds with $v_2=v_3=0$.}
  \label{fig:CIR_vs_speeds}
\end{figure}

\begin{figure}[!t]
    \centering

    \includegraphics[width=\columnwidth]{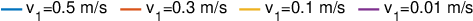}

    \subfloat[CT magnitude.\label{fig:ct_mag}]{%
        \includegraphics[width=0.48\columnwidth]{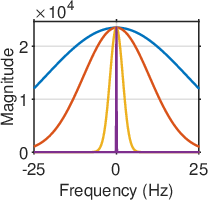}%
    }\hfill
    \subfloat[DT magnitude.\label{fig:dt_mag}]{%
        \includegraphics[width=0.48\columnwidth]{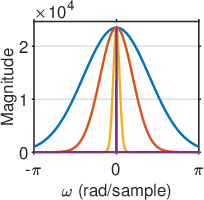}%
    }


    \subfloat[CT phase.\label{fig:ct_phase}]{%
        \includegraphics[width=0.48\columnwidth]{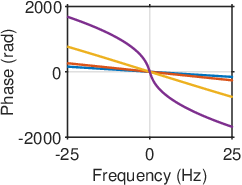}%
    }\hfill
    \subfloat[DT phase.\label{fig:dt_phase}]{%
        \includegraphics[width=0.48\columnwidth]{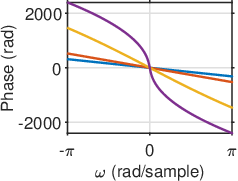}%
    }

    \caption{Continuous-time (CT) and approximate discrete-time (DT) frequency responses of the channel for different constant flow speeds, with $v_2=v_3=0$.}
    \label{fig:ct_dt_frequency_responses}
\end{figure}

The constant-flow CIR is plotted for different flow speeds with $v_2=v_3=0$, using a point source at $\mathbf{z}_0=[0\,0\,0]~\mathrm{m}$, a detection point at $\mathbf{x}=[0.5\,0\,0]~\mathrm{m}$, and $D=6.7698\times10^{-6}~\mathrm{m^2/s}$. The results are shown in \autoref{fig:CIR_vs_speeds}. As the flow speed decreases, the CIR peak occurs later and the response becomes more dispersed in time. For the same configuration, \autoref{fig:ct_dt_frequency_responses} shows the continuous-time and approximate discrete-time magnitude and phase responses, with $f_s=100~\mathrm{Hz}$. Stronger flow yields a larger channel bandwidth and an approximately linear phase response over the plotted frequency range, whereas weaker flow produces more pronounced phase nonlinearity.

\section{Discrete-Time Communication Model}

\subsection{End-to-End Signal Representation}

\begin{figure*}[t]
    \centering
    \includegraphics[width=\textwidth]{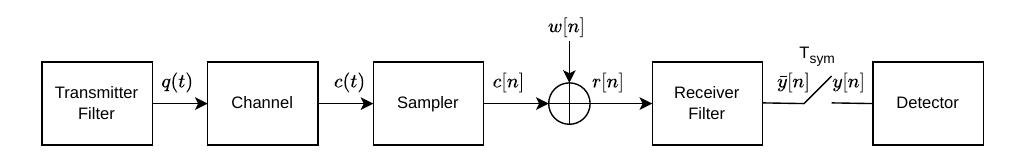}
    \caption{Particle communication link with transmit pulse shaping, diffusion-advection propagation, concentration sampling, additive receiver noise, receiver filtering, and symbol-rate detection.}
    \label{Comm_system_diagram_V1}
\end{figure*}

Building on the discrete-time channel derived above, we formulate the complete communication link. We focus on single-pulse transmissions, where each symbol $s_m$ scales a common transmit pulse $p[n]$, yielding $q[n]=\sum_{m=0}^{N_{\mathrm{sym}}-1}s_m p[n-mN_p]$. The overall system diagram is shown in \autoref{Comm_system_diagram_V1}.

Defining the effective pulse after the physical channel as $g[n]\triangleq h[n]*p[n]$, the received signal before receiver filtering is written as
\begin{equation}
\begin{aligned}
r[n]
&=h[n]*q[n]+w[n]=\sum_{m=0}^{N_{\mathrm{sym}}-1}s_m g[n-mN_p]+w[n],
\end{aligned}
\label{eq_received_signal_before_filter}
\end{equation}
\noindent where \(w[n]\) denotes the noise. Let \(p_{\mathrm{rx}}[n]\) denote the receiver filter, which also contains \(N_p\) coefficients. The receiver-filter output is $\bar y[n]
=r[n]*p_{\mathrm{rx}}[n]=\sum_{m=0}^{N_{\mathrm{sym}}-1}s_m x[n-mN_p]+\bar v[n]$, where $x[n]\triangleq p_{\mathrm{rx}}[n]*g[n]$ and $\bar v[n]=w[n]*p_{\mathrm{rx}}[n]$. The receiver samples $\bar y[n]$ once every $N_p$ samples, with sampling offset $n_0\in\mathbb Z_{\ge0}$ chosen to account for the propagation and receiver-filter delays. The symbol-rate output is $y[i]\triangleq\bar y[n_0+iN_p],i=0,\ldots,N_{\mathrm{sym}}-1$. Defining $x_\ell\triangleq x[n_0+\ell N_p]$ and $v[i]\triangleq \bar v[n_0+iN_p]$, the symbol-rate input-output relation is written as
\begin{equation}
y_i=y[i]
=
\sum_{m=0}^{N_{\mathrm{sym}}-1}
s_m x_{i-m}
+
v[i].
\label{eq_scalar_symbol_rate_model}
\end{equation}
The transmit-pulse vector is defined as $\mathbf p\triangleq
\begin{bmatrix}
p[0] &
p[1] &
\cdots &
p[N_p-1]
\end{bmatrix}^{T}$ and the length-$L_h$ discrete-time channel vector, obtained by truncating the channel tail, is defined as $\mathbf h\triangleq
\begin{bmatrix}
h[0] &
h[1] &
\cdots &
h[L_h-1]
\end{bmatrix}^{T}$. Since \(g[n]=h[n]*p[n]\), the length of the full convolution is $L_g=L_h+N_p-1$. $\mathbf g\triangleq
\begin{bmatrix}
g[0] &
\cdots &
g[L_g-1]
\end{bmatrix}^{T}$ is defined and the corresponding convolution can be written as $\mathbf g=\mathbf H\mathbf p$, where $\mathbf H \in \mathbb R^{L_g\times N_p}$ with entries $[\mathbf H]_{n+1,k+1}=h[n-k]$ for $n=0,\ldots,L_h+N_p-2, \, k=0,\ldots,N_p-1$, where $h[\ell]=0$ for $\ell<0$ or $\ell>L_h-1$, i.e., all channel coefficients whose indices are outside \(0,\ldots,L_h-1\) are understood to be zero. Similarly, the symbol vector is defined as $\mathbf s\triangleq
\begin{bmatrix}
s_0 &
\cdots &
s_{N_{\mathrm{sym}}-1}
\end{bmatrix}^{T}$. The received vector before the receiver filtering has length $L_r=(N_{\mathrm{sym}}-1)N_p+L_g=N_{\mathrm{sym}}N_p+L_h-1$. Therefore, $\mathbf r\triangleq
\begin{bmatrix}
r[0] &
\cdots &
r[L_r-1]
\end{bmatrix}^{T}$ is defined. Moreover, $\mathbf G\in\mathbb R^{L_r\times N_{\mathrm{sym}}}$ is defined with entries $[\mathbf G]_{n+1,m+1}=g[n-mN_p]$ for $n=0,\ldots,L_r-1, \, m=0,\ldots,N_{\mathrm{sym}}-1$ where \(g[\ell]=0\) outside \(0\leq\ell\leq L_g-1\). Hence $\mathbf r=\mathbf G\mathbf s+\mathbf w$ is obtained where $\mathbf w\triangleq
\begin{bmatrix}
w[0] &
\cdots &
w[L_r-1]
\end{bmatrix}^{T}$. The receiver-filter vector is written as $\mathbf p_{\mathrm{rx}}\triangleq
\begin{bmatrix}
p_{\mathrm{rx}}[0] &
\cdots &
p_{\mathrm{rx}}[N_p-1]
\end{bmatrix}^{T}$. Therefore, the length of the full receiver-filter output is $L_{\bar y}=L_r+N_p-1=(N_{\mathrm{sym}}+1)N_p+L_h-2$. $\mathbf P_{\mathrm{rx}} \in \mathbb R^{L_{\bar y}\times L_r}$ denotes the Toeplitz convolution matrix generated by \(\mathbf p_{\mathrm{rx}}\). Its entries are $[\mathbf P_{\mathrm{rx}}]_{n+1,k+1}=p_{\mathrm{rx}}[n-k]$, where $p_{\mathrm{rx}}[\ell]=0$ for $\ell<0$ or $\ell>N_p-1$. Therefore, $\bar{\mathbf y}\triangleq
\begin{bmatrix}
\bar y[0] &
\cdots &
\bar y[L_{\bar y}-1]
\end{bmatrix}^{T}$ can be obtained from $\bar{\mathbf y}=\mathbf P_{\mathrm{rx}}\mathbf r=\mathbf P_{\mathrm{rx}}\mathbf G\mathbf s+\bar{\mathbf v}$, where $\bar{\mathbf v}=\mathbf P_{\mathrm{rx}}\mathbf w$. The downsampling operation is represented using the selection matrix $\mathbf D\in\mathbb R^{N_{\mathrm{sym}}\times L_{\bar y}}$ where
\begin{equation}
[\mathbf D]_{i+1,n+1}
=
\begin{cases}
1,
&
n=n_0+iN_p,
\\
0,
&
\text{otherwise},
\end{cases}
\label{eq_D_entries}
\end{equation}
\noindent for \(i=0,\ldots,N_{\mathrm{sym}}-1\). The symbol-rate output vector is $\mathbf y\triangleq
\begin{bmatrix}
y[0] &
\cdots &
y[N_{\mathrm{sym}}-1]
\end{bmatrix}^{T}$, and satisfies $\mathbf y=\mathbf D\bar{\mathbf y}$. Thus, $\mathbf y=\mathbf D\mathbf P_{\mathrm{rx}}\mathbf G\mathbf s+\mathbf v$ is obtained where $\mathbf v=\mathbf D\bar{\mathbf v}=\mathbf D\mathbf P_{\mathrm{rx}}\mathbf w$. The overall pulse $x[n]$ defined above has length $L_x=L_g+N_p-1=L_h+2N_p-2$, therefore $\mathbf x\triangleq
\begin{bmatrix}
x[0]&
\cdots&
x[L_x-1]
\end{bmatrix}^{T}$ is written. Before downsampling, the effective matrix is defined as $\bar{\mathbf X}\triangleq\mathbf P_{\mathrm{rx}}\mathbf G$ with entries $[\bar{\mathbf X}]_{n+1,m+1}=x[n-mN_p]$, where \(x[\ell]=0\) outside \(0\leq\ell\leq L_x-1\). Hence, $\bar{\mathbf y}=\bar{\mathbf X}\mathbf s+\bar{\mathbf v}$ is obtained. After downsampling, we define $\mathbf X\triangleq\mathbf D\bar{\mathbf X}=\mathbf D\mathbf P_{\mathrm{rx}}\mathbf G$ with entries $[\mathbf X]_{i+1,m+1}=[\bar{\mathbf X}]_{n_0+iN_p+1,m+1}=x[n_0+(i-m)N_p]=x_{i-m}$ for $i,m=0,1,\ldots,N_{\mathrm{sym}}-1$. The complete symbol-rate model can be written as $\mathbf y=\mathbf X\mathbf s+\mathbf v$.


The zero-mean WSS white Gaussian noise $w[n]$ has autocovariance $\mathbb E[w[n_1]w[n_2]]=\sigma_w^2\delta[n_1-n_2]$. The receiver-filtered noise term is $\bar v[n]=\sum_{a=0}^{N_{\rm p}-1}p_{\rm rx}[a]w[n-a]$ and its autocorrelation is $R_{\bar v}[m]=\mathbb E[\bar v[n]\bar v[n-m]]=\sigma_w^2\sum_{a=0}^{N_{\rm p}-1}p_{\rm rx}[a]p_{\rm rx}[a-m]$. Hence, the receiver-filtered noise \(\bar v[n]\) is generally colored. The sampled receiver-output noise is $v_i=v[i]=\bar v[n_0+iN_p]$, whose autocorrelation is $R_v[q]=\mathbb E[v_i v_{i-q}]=\sigma_w^2\sum_{a=0}^{N_{\rm p}-1}p_{\rm rx}[a]p_{\rm rx}[a-qN_p]=\sigma_w^2 \|\mathbf p_{\mathrm{rx}}\|_2^2 \delta[q]$. Therefore, the symbol noise is also WSS white Gaussian.

\subsection{Pulse Design for Particle Communication}

Classical zero-intersymbol interference (ISI) pulse design is based on frequency-domain criteria \cite{Digital_Communications_Proakis}. Particle communication additionally requires nonnegative transmit signals. We therefore formulate the pulse-design problem in the time domain. The scalar relation in \eqref{eq_scalar_symbol_rate_model} can be written as
\begin{equation}
\begin{aligned}
y[i]
&=
\sum_{\ell=i-(N_{\mathrm{sym}}-1)}^{i}
x_{\ell}s_{i-\ell}
+
v[i].
\end{aligned}
\label{eq_desired_and_isi_scalar}
\end{equation}
The desired contribution is $x_0s_i$, whereas the remaining summation represent ISI. For each $\ell=-(N_{\mathrm{sym}}-1),\ldots,N_{\mathrm{sym}}-1$ we define $\mathbf H_\ell\in\mathbb R^{N_p\times N_p}$ by $[\mathbf H_\ell]_{a+1,b+1}=h[n_0+\ell N_p-a-b]$, where $a,b=0,1,\ldots,N_p-1$. Then, $x_\ell=x[n_0+\ell N_p]$ can be written as $x_\ell=\sum_{a=0}^{N_p-1}\sum_{b=0}^{N_p-1}p_{\mathrm{rx}}[a]\,h[n_0+\ell N_p-a-b]\,p[b]=\sum_{a=0}^{N_p-1}\sum_{b=0}^{N_p-1}p_{\mathrm{rx}}[a]\,[\mathbf H_\ell]_{a+1,b+1}\,p[b]$ which becomes $x_\ell=\mathbf p_{\mathrm{rx}}^T\mathbf H_\ell\mathbf p$. Matched-filter detection is optimal for hypothesis testing involving a known deterministic signal in AWGN \cite{Fundamentals_of_Statistical_Signal_Processing_Volume_2}. Motivated by this result, we choose the receiver filter as the time-reversed transmit pulse, while constraining the transmit pulse to be nonnegative, $p[n]\ge0$. Moreover, the pulse is set to unit energy, i.e., $\mathbf p^T\mathbf p=1$. Time reversal is represented by the matrix $\mathbf J\in\mathbb R^{N_p\times N_p}$, defined as
\begin{equation}
[\mathbf J]_{a+1,b+1}
=
\begin{cases}
1,
&
a+b=N_p -1,
\\
0,
&
\text{otherwise},
\end{cases}
\label{eq_J_entries}
\end{equation}
\noindent Thus, $\mathbf p_{\mathrm{rx}}=\mathbf J\mathbf p$. Therefore, $x_\ell=\mathbf p_{\mathrm{rx}}^T\mathbf H_\ell\mathbf p = (\mathbf J\mathbf p)^T\mathbf H_{\ell}\mathbf p = \mathbf p^T\mathbf R_{\ell}\mathbf p$, where $\mathbf R_{\ell}=\mathbf J\mathbf H_{\ell}$. The effective coefficients $x_\ell$ can be found in this way. For a packet of $N_{\rm sym}$ symbols, the packet-level desired contribution is proportional to $N_{\rm sym}x_0^2$. For a fixed symbol lag $\ell$, the coefficient \(x_\ell\) does not appear in every received sample. It appears in the \(i\)-th received sample through the term \(x_\ell s_{i-\ell}\), and this term is present only when \(s_{i-\ell}\) is one of the transmitted packet symbols. Hence, while \(i\) ranges over \(0,\ldots,N_{\mathrm{sym}}-1\), the additional condition \(0\le i-\ell\le N_{\mathrm{sym}}-1\) must also hold. For \(\ell\ge0\), this gives \(i=\ell,\ldots,N_{\mathrm{sym}}-1\), resulting in \(N_{\mathrm{sym}}-\ell\) appearances. For \(\ell<0\), it gives \(N_{\mathrm{sym}}-|\ell|\) appearances. Thus, in both cases, the coefficient \(x_\ell\) contributes \(N_{\mathrm{sym}}-|\ell|\) times over the packet. Therefore, the ISI power term is $\sum_{\ell=-(N_{\mathrm{sym}}-1),\, \ell\neq 0}^{N_{\mathrm{sym}}-1} (N_{\mathrm{sym}}-|\ell|)x_{\ell}^2$. The pulse design minimizes the inverse SINR objective which is written as
\begin{equation}
J(\mathbf p)
=
\frac{
\displaystyle
\sum_{\substack{\ell=-(N_{\rm sym}-1)\\ \ell\neq 0}}^{N_{\rm sym}-1}
(N_{\rm sym}-|\ell|)
\left(\mathbf p^T\mathbf R_{\ell}\mathbf p\right)^2
+
N_{\rm sym}\sigma_w^2 \mathbf p^T\mathbf p
}{
N_{\rm sym}
\left(\mathbf p^T\mathbf R_0\mathbf p\right)^2
}.
\end{equation}
\noindent Therefore, the optimization problem is written as
\begin{equation}
\label{Optimization_Problem}
\begin{aligned}
\min_{\mathbf{p}} \quad & J(\mathbf{p}) \\
\text{subject to} \quad & \mathbf{p} \ge 0, \qquad \mathbf{p}^T\mathbf{p} = 1.
\end{aligned}
\end{equation}

\subsection{Detection Methods}

We employ detection methods that are well established in digital communications and briefly summarize their mathematical operations as a reference for their application within the proposed particle communication framework. These include symbol-by-symbol detection under an ISI-free model, maximum likelihood (ML) sequence detection, linear minimum mean square error (LMMSE) equalization, and decision feedback equalization (DFE).

\subsubsection{Symbol-by-Symbol Detection under an ISI-Free Model}
As a simple detection baseline, we make independent symbol decisions under the ISI-free model $y_i=x_0s_i+v_i$, obtained by retaining only the desired-symbol contribution in \eqref{eq_desired_and_isi_scalar}. Since $v_i\sim\mathcal N(0,\sigma_w^2)$, the conditional likelihood is $p(y_i|s_i=a)=\frac{1}{\sqrt{2\pi\sigma_w^2}}\exp[-\frac{(y_i-x_0a)^2}{2\sigma_w^2}]$ for $a\in\mathcal A$, where \(\mathcal A\) is the modulation alphabet. Maximizing this likelihood gives the symbol-by-symbol decision rule $\widehat s_i=\arg\min_{a\in\mathcal A}(y_i-x_0a)^2$. For $x_0>0$, this is equivalent to mapping $y_i/x_0$ to the nearest constellation point. Although the actual received signal contains ISI, this detector does not compensate for it. Its performance therefore provides a baseline for assessing the benefits of pulse shaping and the ISI-aware detection methods described below.

\subsubsection{Maximum Likelihood Sequence Detection}
The finite-window detector model is written as $\mathbf y=\widehat{\mathbf X}\mathbf s+\mathbf v$ where \(\widehat{\mathbf X}\) is the finite-window ISI matrix used by the detector. Its entries are
\begin{equation}
[\widehat{\mathbf X}]_{i+1,m+1}
=
\begin{cases}
x_{i-m}, & -L_-\le i-m\le L_+,\\
0, & \text{otherwise},
\end{cases}
\end{equation}
\noindent for $i,m=0,1,\ldots,N_{\rm sym}-1$. The parameters \(L_-\) and \(L_+\) define the retained precursor and postcursor ISI ranges, respectively. Since $\mathbf v\sim\mathcal N(\mathbf 0,\sigma_w^2\mathbf I_{N_{\rm sym}})$, the conditional distribution of \(\mathbf y\) given \(\mathbf s\) is $\mathbf y|\mathbf s\sim\mathcal N(\widehat{\mathbf X}\mathbf s,\sigma_w^2\mathbf I_{N_{\rm sym}})$. The corresponding conditional probability density is
\begin{equation}
p(\mathbf y|\mathbf s)
=
\frac{1}{(2\pi\sigma_w^2)^{N_{\rm sym}/2}}
\exp
\left[
-\frac{1}{2\sigma_w^2}
\|\mathbf y-\widehat{\mathbf X}\mathbf s\|_2^2
\right].
\end{equation}
The ML sequence detector selects the sequence maximizing this likelihood, $\widehat{\mathbf s}_{\rm ML}=\arg\max_{\mathbf s\in\mathcal A^{N_{\rm sym}}}p(\mathbf y|\mathbf s)$. This is equivalent to $\widehat{\mathbf s}_{\rm ML}=\arg\min_{\mathbf s\in\mathcal A^{N_{\rm sym}}}\|\mathbf y-\widehat{\mathbf X}\mathbf s\|_2^2$. It is written as $\widehat{\mathbf s}_{\rm ML}=\arg\min_{\mathbf s\in\mathcal A^{N_{\rm sym}}}\sum_{i=0}^{N_{\rm sym}-1}(y_i-\sum_{\ell=-L_-}^{L_+}x_\ell s_{i-\ell})^2$ in scalar form. Direct exhaustive ML detection is generally infeasible because it requires testing \(|\mathcal A|^{N_{\rm sym}}\) candidate sequences. Therefore, the Viterbi algorithm is used to compute the same finite-window ML solution when the retained memory is sufficiently short \cite{MaximumLikelihood_Sequence_Estimation_of_Digital_Sequences_in_the_Presence_of_Intersymbol_Interference, How_I_Learned_to_Love_the_Trellis}. The retained ISI memory is $M \triangleq L_-+L_+$. At trellis time \(t\), the state is chosen as the previous \(M\) symbols, $\mathbf q_{t-1}=[s_{t-M},s_{t-M+1},\ldots,s_{t-1}]$. A trellis branch appends a candidate symbol \(s_t\in\mathcal A\), producing the new state $\mathbf q_t=[s_{t-M+1},s_{t-M+2},\ldots,s_t]$. Because the model contains precursor ISI, \(y_i\) depends on future symbols up to \(s_{i+L_-}\). Therefore, \(y_i\) can only be scored after \(s_{i+L_-}\) has been hypothesized. At trellis time \(t\), the newest hypothesized symbol is \(s_t\), so the observation scored at that branch is $i=t-L_-$. The branch metric is therefore $\Gamma_t=(y_{t-L_-}-\sum_{\ell=-L_-}^{L_+}x_\ell s_{t-L_- - \ell})^2$. The accumulated path metric is updated as $P_t^{\rm cand}(\mathbf q_t)=P_{t-1}(\mathbf q_{t-1})+\Gamma_t$. If multiple branches enter the same state, only the branch with the smallest accumulated path metric is retained. The initial trellis state is determined by the last $M$ known pilot symbols immediately preceding the data sequence, rather than assumed to be an all-zero state. Consequently, at least $M$ pilot symbols are required. The trellis contains $|\mathcal A|^M=|\mathcal A|^{L_-+L_+}$ states, making exact Viterbi MLSE practical only for sufficiently short retained memory. For long-memory channels, MLSE provides a theoretical benchmark, while LMMSE equalization and DFE offer computationally more tractable alternatives.

\subsubsection{Linear MMSE Equalization}
The nonnegative symbol amplitudes have a nonzero mean, which must be accounted for in LMMSE estimation. We have a linear observation model $\mathbf y=\mathbf X\mathbf s+\mathbf v$. We seek an affine estimator $\hat{\mathbf{s}}=\mathbf a+\mathbf B\mathbf y$ by solving $\arg\min_{\mathbf a,\mathbf B}\mathbb E[\|\mathbf s-\hat{\mathbf{s}}\|_2^2]$. For fixed $\mathbf B$, the minimizing $\mathbf a$ can be found as $\mathbf a = E[\mathbf{s}]-\mathbf B E[\mathbf y]$. The means are $E[\mathbf{s}] \triangleq \mathbf m_s$ and $E[\mathbf y] \triangleq \mathbf m_y = \mathbf X \mathbf m_s$. Defining $\tilde{\mathbf{s}} \triangleq \mathbf{s}-\mathbf m_s$ and $\tilde{\mathbf{y}} \triangleq \mathbf{y}-\mathbf m_y$, the minimization problem reduces to $\arg\min_{\mathbf B} E[||\tilde{\mathbf{s}} - \mathbf B \tilde{\mathbf{y}}||_2^2]$. The solution yields $\mathbf B = \mathbf C_{sy} \mathbf C_{yy}^{-1}$ \cite{Fundamentals_of_statistical_signal_processing_estimation_theory}. Assuming the symbols and the receiver noise are independent, we have $\mathbf C_{sy} \triangleq E[\tilde{\mathbf{s}}\tilde{\mathbf{y}}^{T}] = \mathbf C_{s} \mathbf X^{T}$ and $\mathbf C_{yy} \triangleq E[\tilde{\mathbf{y}}\tilde{\mathbf{y}}^{T}] =  \mathbf X \mathbf C_{s} \mathbf X^{T} + \mathbf C_{v}$ where $\mathbf C_s$ and $\mathbf C_v$ are the symbol and noise covariance matrices, respectively. The resulting LMMSE estimator is $\hat{\mathbf{s}} =  \mathbf m_s + \mathbf C_{s}  \mathbf X^{T}(\mathbf X \mathbf C_{s} \mathbf X^{T}+\mathbf C_{v})^{-1}(\mathbf y - \mathbf X \mathbf m_s)$. For i.i.d. symbols with $\mathbf C_{s} = \sigma_s^2\mathbf I$ and $\mathbf C_{v} = \sigma_w^2\mathbf I$, this result simplifies to $\hat{\mathbf{s}}=(\mathbf X^{T}\mathbf X+\lambda \mathbf I)^{-1}(\mathbf X^{T}\mathbf{y}+\lambda \mathbf m_s)$, where $\lambda = \frac{\sigma_w^2}{\sigma_s^2}$. The resulting symbol estimates are mapped to the nearest constellation points.

\subsubsection{Decision Feedback Equalization}
To account for the nonzero symbol mean, we define the centered symbol $a_i\triangleq s_i-\mu_s$, where $\mu_s$ denotes the mean of the transmitted symbols. Moreover, we define the mean-removed observation $z_i \triangleq y_i-\mu_s\sum_{\ell=i-(N_{\mathrm{sym}}-1)}^{i}x_{\ell}$. Substituting these definitions into \eqref{eq_desired_and_isi_scalar} yields $z_i=\sum_{\ell=i-(N_{\mathrm{sym}}-1)}^{i}x_{\ell}a_{i-\ell}+v_i$. The DFE output for estimating \(a_i\) is \cite{Digital_Communications_Proakis} $\widehat a_i=\sum_{j=-K_1}^{0}c_j z_{i-j}+\sum_{r=1}^{K_2}c_r\widetilde a_{i-r}$, where \(c_{-K_1},\ldots,c_0\) are feedforward coefficients, \(c_1,\ldots,c_{K_2}\) are feedback coefficients, and \(\widetilde a_{i-r}\) denotes a previous decision on a centered symbol. Assuming correct past decisions and substituting the expression for $z_i$ into the feedforward section yields $\widehat a_i=\sum_{r=i-(N_{\mathrm{sym}}-1)}^{i}q_r a_{i-r}+\sum_{j=-K_1}^{0}c_jv_{i-j}+\sum_{r=1}^{K_2}c_r a_{i-r}$, where $q_r=\sum_{j=-K_1}^{0}c_jx_{r-j}$. The coefficient multiplying the past symbol \(a_{i-r}\), for \(r=1,\ldots,K_2\), is \(q_r+c_r\). Therefore, the feedback coefficients are chosen as $c_r=-q_r, \, r=1,\ldots,K_2$. Thus, the feedback coefficients are determined from the feedforward coefficients by $c_r=-\sum_{j=-K_1}^{0}c_jx_{r-j},\, r=1,\ldots,K_2$. Assuming that the feedback section cancels the considered postcursor terms, the DFE output becomes $\widehat a_i=q_0a_i+\sum_{r=i-(N_{\mathrm{sym}}-1)}^{-1}q_ra_{i-r}+\sum_{j=-K_1}^{0}c_jv_{i-j}$. The equalizer coefficients are obtained using the MSE criterion \cite{Optimum_MeanSquare_Decision_Feedback_Equalization}. The corresponding estimation error is $e_i \triangleq a_i-\widehat a_i$. Using $E[a_ia_m]=\sigma_a^2\delta_{i,m}$, the DFE mean-square error is found as $J_i=\sigma_a^2(1-2q_0+\sum_{r=i-(N_{\mathrm{sym}}-1)}^{0}|q_r|^2)+\sigma_w^2\sum_{j=-K_1}^{0}|c_j|^2$. The feedforward taps are obtained by minimizing \(J_i\) with respect to \(c_{-K_1},\ldots,c_0\). This corresponds to solving the linear system $\boldsymbol{\Psi}\mathbf c_{\mathrm{FF}}=\mathbf b$, where $[\boldsymbol{\Psi}]_{p,j} \triangleq \sigma_a^2 \sum_{r=i-(N_{\mathrm{sym}}-1)}^{0} x_{r-j}x_{r-p}+\sigma_w^2\delta_{p,j}$, $[\mathbf b]_p\triangleq \sigma_a^2x_{-p}$, and $\mathbf c_{\mathrm{FF}} \triangleq \begin{bmatrix} c_{-K_1} & c_{-K_1+1} & \cdots & c_0 \end{bmatrix}^{T}$. The shifted estimate $\widehat a_i+\mu_s$ is mapped to the nearest constellation point to obtain the symbol decision $\widehat s_i$. The centered decision used in the feedback path is then $\widetilde a_i=\widehat s_i-\mu_s$.

\section{Simulation Results}

All numerical experiments are performed in MATLAB R2023b using macroscopic channel parameters. We set $D=6.7698\times10^{-6}~\mathrm{m^2/s}$, place the transmitter at $\mathbf z_0=[0\,0\,0]~\mathrm{m}$, and direct the constant flow toward the receiver along the positive $x_1$ axis, with $v_2=v_3=0$. The analytical approximation validation uses a transmitter-receiver distance of $R=0.5~\mathrm{m}$, whereas the pulse-design and BER experiments use $R=1~\mathrm{m}$. The digital sampling frequency is $f_s=100~\mathrm{Hz}$.

\begin{figure}[!h]
    \centering
    \subfloat[]{%
        \includegraphics[width=0.48\columnwidth]{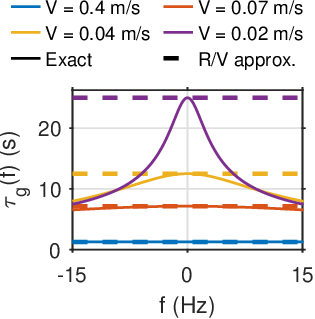}%
        \label{fig:group_delay_validation}%
    }\hfill
    \subfloat[]{%
        \includegraphics[width=0.48\columnwidth]{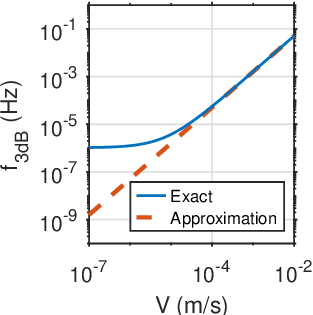}%
        \label{fig:bandwidth_validation}%
    }
    \caption{Validation of the advection-dominated approximations for
    \protect\subref{fig:group_delay_validation} group delay and
    \protect\subref{fig:bandwidth_validation} one-sided $3$-dB bandwidth.}
    \label{fig:delay_bandwidth_validation}
\end{figure}

\autoref{fig:delay_bandwidth_validation}\protect\subref{fig:group_delay_validation} compares the exact group delay in \eqref{eq:exact_group_delay} with the approximation in \eqref{eq:constant_group_delay_approximation} for $V=0.4$, $0.07$, $0.04$, and $0.02~\mathrm{m/s}$. The approximation agrees closely with the exact response for the strongest flow over the plotted frequency range. As the flow speed decreases, the exact group delay becomes increasingly frequency dependent, and the constant approximation departs from it away from zero frequency. This behavior is consistent with the condition $8\pi D|f|/V^2\ll1$. \autoref{fig:delay_bandwidth_validation}\protect\subref{fig:bandwidth_validation} compares \eqref{eq:three_db_bandwidth} with the bandwidth obtained numerically by solving $|H(f_{\mathrm{3dB}})|/|H(0)|=1/\sqrt{2}$ over $10^{-7}\leq V\leq10^{-2}~\mathrm{m/s}$. The curves converge as the flow speed increases, supporting the approximate $V^{3/2}$ scaling. At very low speeds, the exact bandwidth approaches a diffusion-controlled limit, while the approximation continues to decrease, illustrating its limited applicability outside the advection-dominated regime.

The pulse-design problem in \eqref{Optimization_Problem} is solved numerically using MATLAB's \texttt{fmincon} with the interior-point algorithm and analytical gradients of the objective and unit-energy constraint. Multiple feasible initializations are used, including a rectangular pulse, impulses at the first, middle, and last samples, and random nonnegative unit-energy pulses. The solution with the smallest objective value is retained. The sampling offset is fixed as $n_0=n_{\mathrm{ch}}+n_{\mathrm{pd}}$, where $n_{\mathrm{ch}}$ is a channel-delay reference in samples and $n_{\mathrm{pd}}=N_p-1$ accounts for the time reversal of the length-$N_p$ transmit pulse in the receiver filter. We choose $n_{\mathrm{ch}}$ near the advective transit delay $R/(VT_s)$, using $n_{\mathrm{ch}}=20000$, $3333$, and $1000$ for $V=0.005$, $0.03$, and $0.1~\mathrm{m/s}$, respectively. This offset is used consistently in pulse design and detection. Perfect timing knowledge is assumed throughout the BER experiments; timing acquisition and robustness to synchronization errors remain directions for future research.

\begin{figure}[!h]
    \centering
    \subfloat[]{%
        \includegraphics[width=0.48\columnwidth]{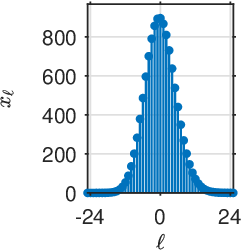}%
        \label{fig:xl_high_memory}%
    }\hfill
    \subfloat[]{%
        \includegraphics[width=0.48\columnwidth]{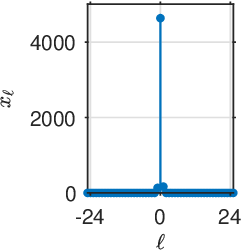}%
        \label{fig:xl_low_memory}%
    }

    \caption{Symbol-rate coefficients $x_\ell$ for the optimized
    $10$-dB pulses in the
    \protect\subref{fig:xl_high_memory} high-memory and
    \protect\subref{fig:xl_low_memory} low-memory channels.}
    \label{fig:symbol_rate_memory}
\end{figure}

\autoref{fig:symbol_rate_memory} shows the effective symbol-rate coefficients $x_\ell$ for the pulses optimized at a nominal SNR of $10~\mathrm{dB}$, with $T_{\mathrm{sym}}=2~\mathrm{s}$, $N_p=200$, and $N_{\mathrm{sym}}=300$. The desired-symbol contribution is represented by $x_0$, while coefficients with $\ell<0$ and $\ell>0$ represent precursor and postcursor ISI, respectively. For $V=0.005~\mathrm{m/s}$, coefficients extend across several neighboring symbols on both sides of $\ell=0$, indicating that both ISI contributions must be considered. For $V=0.03~\mathrm{m/s}$, the coefficients are concentrated more closely around the desired-symbol contribution, with much smaller contributions from neighboring symbols. Accordingly, we refer to the channels with $V=0.005$ and $0.03~\mathrm{m/s}$ as the high-memory and low-memory channels, respectively. This classification describes their effective memory relative to the common symbol duration and is used in the following pulse-design and BER comparisons.

\begin{figure}[h]
    \centering
    \subfloat[]{%
        \includegraphics[width=0.315\columnwidth]{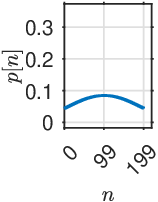}%
        \label{fig:pulse_lm_m20}%
    }\hfill
    \subfloat[]{%
        \includegraphics[width=0.315\columnwidth]{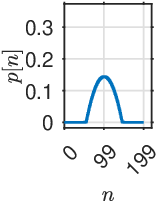}%
        \label{fig:pulse_lm_10}%
    }\hfill
    \subfloat[]{%
        \includegraphics[width=0.315\columnwidth]{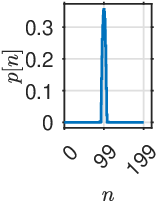}%
        \label{fig:pulse_lm_40}%
    }
    \par\vspace{1mm}
    \subfloat[]{%
        \includegraphics[width=0.315\columnwidth]{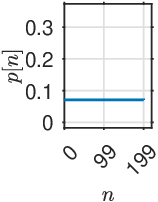}%
        \label{fig:pulse_hm_m20}%
    }\hfill
    \subfloat[]{%
        \includegraphics[width=0.315\columnwidth]{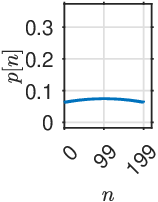}%
        \label{fig:pulse_hm_10}%
    }\hfill
    \subfloat[]{%
        \includegraphics[width=0.315\columnwidth]{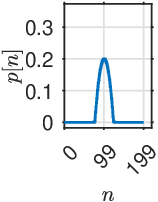}%
        \label{fig:pulse_hm_40}%
    }
    \caption{Optimized pulses for the low-memory
    \protect\subref{fig:pulse_lm_m20}--\protect\subref{fig:pulse_lm_40}
    and high-memory
    \protect\subref{fig:pulse_hm_m20}--\protect\subref{fig:pulse_hm_40}
    channels at $-20$, $10$, and $40$ dB from left to right.}
    \label{fig:optimized_pulses}
\end{figure}

\autoref{fig:optimized_pulses} shows the optimized pulses for these two channel conditions. Separate pulses are designed at noise variances calibrated to nominal SNRs of $-20$, $10$, and $40~\mathrm{dB}$. At low SNR, the pulses occupy a broad portion of the symbol interval, with the high-memory channel producing an almost rectangular pulse. As SNR increases, the pulse optimized for the low-memory channel becomes progressively narrower and approaches an impulse-like shape. The pulse optimized for the high-memory channel also concentrates at high SNR but remains broader. These shapes reflect the balance between desired-signal gain and ISI-noise in the design objective: broader pulses are favored when noise is substantial, whereas temporal concentration becomes more beneficial as noise decreases.

The BER experiments compare the optimized pulse with a full-duration rectangular pulse and a single-sample impulse placed at the middle of the symbol interval. All pulses have unit energy, and each receiver filter is the time-reversed version of its corresponding transmit pulse. Packets contain $N_{\mathrm{sym}}=300$ symbols, including $20$ known pilots, with equiprobable binary data symbols from $\mathcal A=\{1,2\}$. Silent tail intervals are appended to retain the delayed received waveform. SNR is calculated before receiver filtering. The noise variance is adjusted for each pulse and operating point to achieve the requested SNR, and the optimized pulse is designed separately for the corresponding channel and noise level. BER is calculated by dividing the total data-bit errors by the total tested data bits across Monte Carlo trials. Colors identify the pulses and line styles identify the detectors. Error bars show nominal $95\%$ binomial confidence intervals; downward triangles mark their upper endpoints at operating points with zero observed errors.

\begin{figure}[!t]
    \centering
    \includegraphics[width=\columnwidth]{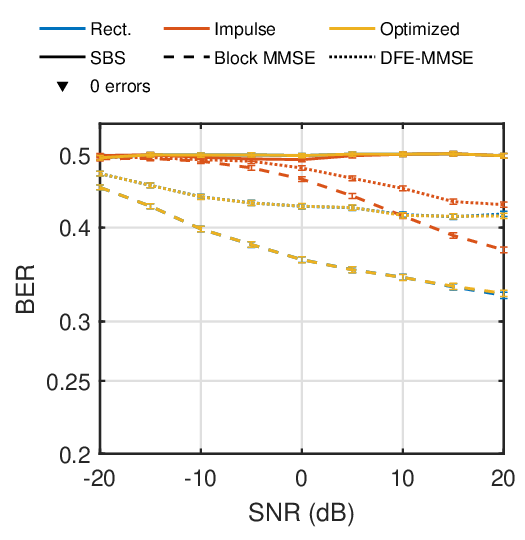}
    \caption{BER versus SNR for $v_1=0.005\,\mathrm{m/s}$ and
    $T_{\mathrm{sym}}=2\,\mathrm{s}$, aggregated over 300
    Monte Carlo trials.}
    \label{fig:ber_snr_high_memory}
\end{figure}

\begin{figure}[!h]
    \centering
    \includegraphics[width=\columnwidth]{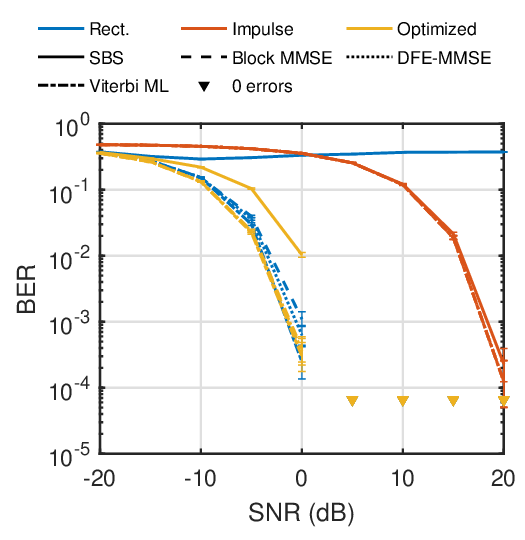}
    \caption{BER versus SNR for $v_1=0.03\,\mathrm{m/s}$ and
    $T_{\mathrm{sym}}=2\,\mathrm{s}$, aggregated over 201
    Monte Carlo trials.}
    \label{fig:ber_snr_low_memory}
\end{figure}

\autoref{fig:ber_snr_high_memory} and \autoref{fig:ber_snr_low_memory} show BER versus SNR from $-20$ to $20~\mathrm{dB}$ in $5$-dB steps, with $T_{\mathrm{sym}}=2~\mathrm{s}$. Symbol-by-symbol detection, block LMMSE equalization, and DFE are evaluated in both channels. Viterbi MLSE is additionally evaluated in the low-memory channel; its state-space requirements are prohibitive for the high-memory case. In the high-memory channel, symbol-by-symbol detection remains close to a BER of $0.5$, indicating that increasing SNR alone cannot overcome the unmodeled ISI. Block LMMSE provides the largest improvement among the evaluated detectors, while DFE produces a smaller reduction. Differences between pulse types are comparatively modest, and substantial errors remain throughout the sweep. In the low-memory channel, the optimized pulse yields a pronounced improvement over the rectangular pulse with symbol-by-symbol detection. For example, at $0~\mathrm{dB}$ its BER is approximately $10^{-2}$, whereas the rectangular pulse remains near $0.4$. With ISI-aware detection, rectangular and optimized pulses achieve much lower BERs and perform closely over much of the sweep, while the single-sample impulse requires higher SNR. Thus, the benefit of pulse design depends on receiver processing.

\begin{figure}[!t]
    \centering
    \includegraphics[width=\columnwidth]{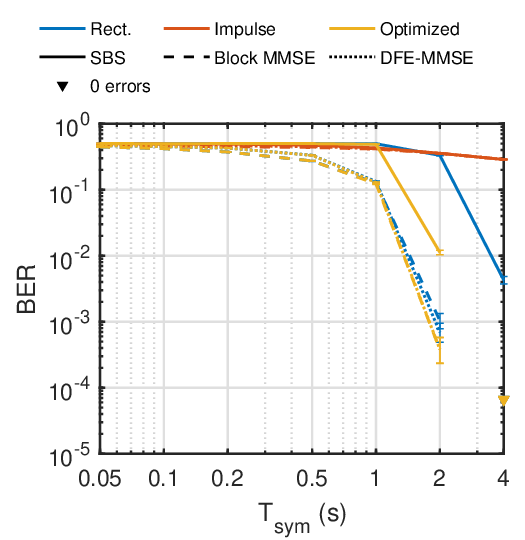}
    \caption{BER versus symbol duration for $v_1=0.03\,\mathrm{m/s}$
    at $\mathrm{SNR}=0\,\mathrm{dB}$, aggregated over 200
    Monte Carlo trials.}
    \label{fig:ber_tsym_flow_003}
\end{figure}

\begin{figure}[!t]
    \centering
    \includegraphics[width=\columnwidth]{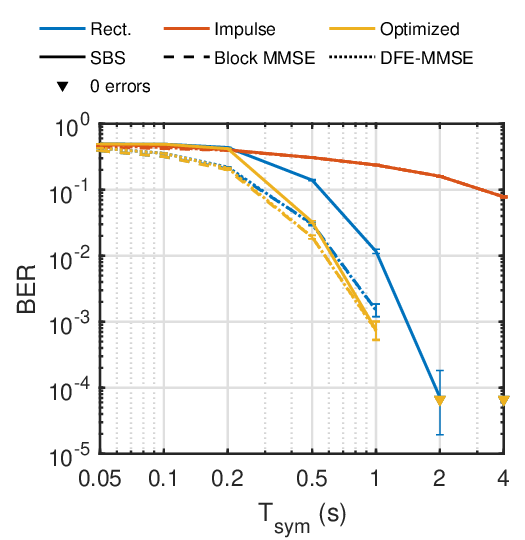}
    \caption{BER versus symbol duration for $v_1=0.1\,\mathrm{m/s}$
    at $\mathrm{SNR}=0\,\mathrm{dB}$, aggregated over 200
    Monte Carlo trials.}
    \label{fig:ber_tsym_flow_01}
\end{figure}

Finally, \autoref{fig:ber_tsym_flow_003} and \autoref{fig:ber_tsym_flow_01} examine the effect of symbol duration at $0~\mathrm{dB}$ for $V=0.03$ and $0.1~\mathrm{m/s}$, respectively. We use $T_{\mathrm{sym}}\in\{0.05,0.1,0.2,0.5,1,2,4\}~\mathrm{s}$ and redesign the optimized pulse for every channel and symbol duration. The sampling offset is updated through $n_{\mathrm{pd}}=N_p-1$, and the number of silent tail intervals is adjusted to preserve the observation duration. Symbol-by-symbol detection, block LMMSE, and DFE are compared; Viterbi MLSE is omitted because short symbol intervals produce prohibitively large retained memory. The SNR convention remains unchanged. Short symbol intervals produce high BER for all pulse types because of strong overlap between neighboring symbol responses. Increasing the symbol duration reduces this overlap and improves performance, with rectangular and optimized pulses benefiting substantially from equalization. The optimized pulse also improves symbol-by-symbol detection. The stronger-flow channel reaches low BER at shorter symbol durations, consistent with its larger bandwidth and reduced temporal dispersion. The single-sample impulse remains less effective at this SNR, illustrating that temporal concentration alone is insufficient when noise is significant.

\section{Conclusion}

This work developed a digital communication framework for three-dimensional free-space diffusion-advection channels, connecting physical transport, frequency-domain characterization, pulse design, and symbol detection. The analysis established that, under advection-dominated conditions, the group delay approaches $R/V$ and the one-sided $3$-dB bandwidth scales as $V^{3/2}$. Nonnegative, unit-energy pulse design explicitly accounted for both precursor and postcursor ISI. The optimized pulses became broader when noise was substantial and more concentrated as noise decreased, illustrating the balance between desired-signal gain and ISI reduction. BER results showed that pulse shaping can substantially improve symbol-by-symbol detection, while its additional benefit depends on the equalization method and channel memory. Increasing SNR alone was insufficient to overcome severe ISI, whereas stronger flow enabled reliable communication at shorter symbol durations. These findings connect transport parameters to practical signaling and receiver choices.

\bibliographystyle{IEEEtran}
\bibliography{references.bib}

\vspace{11pt}

\begin{IEEEbiography}[{\includegraphics[width=1in,height=1.25in,clip,keepaspectratio]{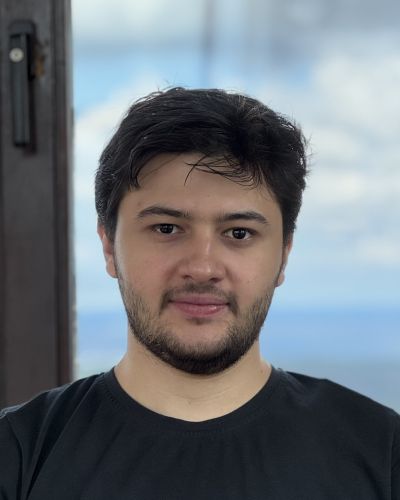}}]{Fatih Merdan}
completed his high school education at Kırıkkale Science High School. He received his B.Sc. degree in Electrical and Electronics Engineering from Middle East Technical University. He is currently pursuing his M.Sc. degree in Electrical and Electronics Engineering under the supervision of Prof. Akan at Koç University, Istanbul, Turkey.
\end{IEEEbiography}

\vspace{11pt}

\begin{IEEEbiography}[{\includegraphics[width=1in,height=1.25in,clip,keepaspectratio]{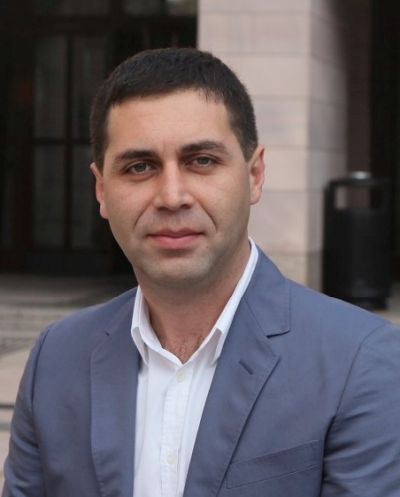}}]{Ozgur B. Akan}
\textbf{(Fellow, IEEE)} received the Ph.D. degree in electrical and computer engineering from
Georgia Institute of Technology, USA, in 2004. He is currently a Professor in electronic and communication engineering with the University of Cambridge, U.K., where he leads the Centre for neXt Communications (CXC). He is also the Director of CXC, Koc¸ University, Turkiye. His research interests include wireless, space, quantum, and molecular communications.
\end{IEEEbiography}

\vfill

\end{document}